\documentclass{elsart41}
\usepackage{graphics}
\usepackage{graphicx}
\usepackage{amssymb}
\begin{document}

\begin{frontmatter}



\title{Point-contact spectroscopy of the normal state excitations in PrOs$_4$Sb$_{12}$}
%

\author[AA]{O. E. Kvitnitskaya\corauthref{Name1}},
\ead{kvitnitskaya@ilt.kharkov.ua}
\author[AA]{Yu. G. Naidyuk},
\author[AA]{I. K. Yanson},\,
\author[BB]{A. Karkin},
\author[BB]{S. Naumov},
\author[BB]{N. Kostromitina}

\address[AA]{B.Verkin Institute for Low Temperature Physics and Engineering, NASU,
47 Lenin Ave., 61103, Kharkiv, Ukraine}
\address[BB]{Institute of Metal Physics, S. Kovalevskoi Str. 18, Ekaterinburg 620219, Russia}

\corauth[Name1]{Corresponding author. Tel: ++380-57-3402211 fax: ++(380)-57-340-33-70}

\begin{abstract}

Point-contact (PC) investigations of the heavy-fermion (HF)
superconductor PrOs$_4$Sb$_{12}$ in the temperature range between
1.5 and 4.2\,K and in a magnetic field up to 8\,T are presented.
The main feature in PC spectra, the second derivative of the I-V
curves, in the normal state is a peak at about 1-3\,mV. This low
energy peak is smeared by the temperature rise and suppressed (or
splitted) by a magnetic field. The origin of the peak is likely
excitation of the Pr$^{3+}$ ion from the ground $\Gamma_1$ state
to the upper $\Gamma_5$ state, considering the generally accepted
crystalline electric field (CEF) schema in PrOs$_4$Sb$_{12}$.
Absence of the visible features of other CEF transitions in the PC
spectra testifies to the dominant coupling of the conducting
electrons with this low-lying (about 1\,meV in energy) excitation.
Thus, this interaction is important for ascertainment of the
nature of HF and superconducting state in PrOs$_4$Sb$_{12}$.

\end{abstract}

\end{frontmatter}

PrOs$_4$Sb$_{12}$ appears to be the first Pr-based heavy-fermion
superconductor (T$_c$= 1.85\,K) with an effective quasiparticle
mass $m^* \approx 50m_e$ according to the enhanced electronic
specific heat coefficient $\gamma$ \cite{Bauer}.
Crystal-electric-field (CEF) splitting of the Pr$^{3+}$ multiplet
influences the physical properties of PrOs$_4$Sb$_{12}$ as widely
accepted, first of all, the heavy-fermion phenomena and
superconducting pairing mechanism in this compound. According to
the recent specific heat, magnetization, elastic and inelastic
neutron scattering measurements \cite{Maple} CEF schema in
PrOs$_4$Sb$_{12}$ tends to be as follows: magnetic $\Gamma_5$ and
$\Gamma_4$ triplets, nonmagnetic $\Gamma_3$ doublet separated from
the ground $\Gamma_1$ state by 10K, 100K and 300K respectively.
Thus, as it is commonly believed, mass enhancement would arise
from inelastic exchange scattering of the conduction electrons by
the well-localized hybridized $4f^2$ electrons of the Pr$^{3+}$
ion.

Point-contact spectroscopy (PCS) turns to be a successful tool for
the direct study of energy-dependent interactions between
conducting electrons and different quasiparticle excitations in
metals \cite{Naidyuk}. In \cite{Omel} a PCS theory for the
$f$-localized states was evolved. It was shown that nonlinear
conductivity of PC is determined by the inelastic scattering of
the conducting electrons on the $f$-shell, what allows to probe
the CEF levels of the rare-earth ion. PCS was used successfully to
study the CEF levels of the Pr$^{3+}$ ion in PrNi$_5$ and their
Zeeman splitting \cite{Reiffers}. In the case of heavy-fermion
materials obtaining of the spectroscopic information is restricted
by their very high resistivity that leads to violation of the
ballistic condition and transition to the thermal regime and
dominance of the self-heating effects in the PC spectra. Only PC`s
on relatively "low resistivity" specimens ($ \leq 10\mu\Omega$cm)
which we believe our PrOs$_4$Sb$_{12}$ samples belong to, may have
at low voltage inelastic electron mean free path  exceeding the PC
size and spectra with spectroscopic features.


We have used the PrOs$_4$Sb$_{12}$ single-crystal samples (T$_c$=
1.75\,K) of a sub-millimeter size. To produce PC`s a Cu electrode
had a gentle touch to `as grown` PrOs$_4$Sb$_{12}$ surface  in
liquid He$^4$. The first and second derivatives of the I-V curves
as function of bias voltage were recorded using a standard lock-in
technique.


In Fig.\,1 the $dV/dI(V)$ characteristics of PC between
PrOs$_4$Sb$_{12}$ and Cu taken at T=1.5\,K are presented for the
magnetic field range between 0 and 8\,T. The `quasi` parabolic
$dV/dI(V)$ curves show also a minimum around V$\simeq$0\,mV.
Magnetic field somewhat degrades the zero-bias minimum.

\begin{figure}[!ht]
\begin{center}
\includegraphics[width=0.5\textwidth]{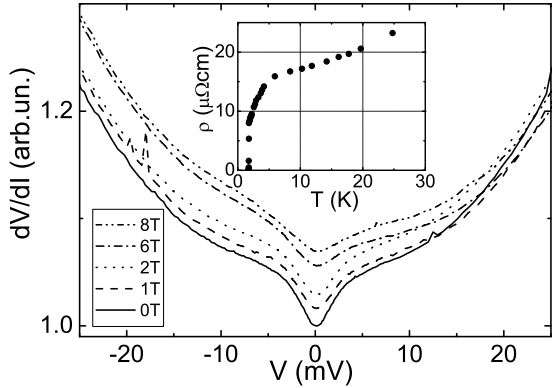}
\end{center}
\caption[] {Experimental differential resistance ($dV/dI$) of a
PrOs$_4$Sb$_{12}$--Cu PC with R$_0$=0.8\,$\Omega$, in magnetic
field at T=1.5\,K. Inset: specific resistivity of
PrOs$_4$Sb$_{12}$. } \label{first}
\end{figure}

Figure\,2 displays PC spectra $d^2V/dI^2(V)$ corresponding to
Fig.\,1. Zero magnetic field spectrum shows sharp peak at about
0.8\,mV, which is suppressed by magnetic field of 0.5\,T. Traces
of superconductivity are likely responsible for this feature.

The most prominent peculiarity of the spectra is a maximum at
about 1.8\,mV. Increasing temperatures (not shown) as well as
magnetic fields lead to its broadening, reduction and finally even
splitting is seen at high magnetic fields. This maximum can be
connected with the allowed CEF transition $\Gamma_1$ $\rightarrow$
$\Gamma_5$ with energy about 0.7\,meV, according to the level
scheme determined in \cite{Goremychkin}. Next allowed CEF
transition (around 11\,meV \cite{Goremychkin}) is not displayed in
the PC spectra pointing out to the dominant coupling of the
conducting electrons with the low-lying $\Gamma_1$ $\rightarrow$
$\Gamma_5$ excitation. Moreover, no features due to phonon maxima
of Cu (around 16\,meV) as well as PrOs$_4$Sb$_{12}$ phonons were
enucleated (top inset in Fig.\,2). This may be associated with the
small contribution from phonon compared to the CEF excitations and
transition of the PC region in the thermal regime at higher
voltages ($>$ 10\,mV) due to decrease of the inelastic mean free
path.

Figure\,2 (bottom inset) shows a calculation according to the CEF
theory \cite{Omel} compared to the experimental curve taken at
1\,T to suppress the mentioned 0.8\,mV peak. The main parameter of
the calculation is the position of the CEF peak taken at 1.1\,meV
to fit the experimental curve what is in a reasonable agreement
with the neutron data \cite{Goremychkin}. Contrary, calculation of
the spectrum in the thermal regime according to Eq.(3.23) in
\cite{Naidyuk} (Fig.\,2, upper inset) exhibits a maximum at
0.6\,mV with a shallow hump around 20\,mV.

\begin{figure}[!ht]
\begin{center}
\includegraphics[width=0.5\textwidth]{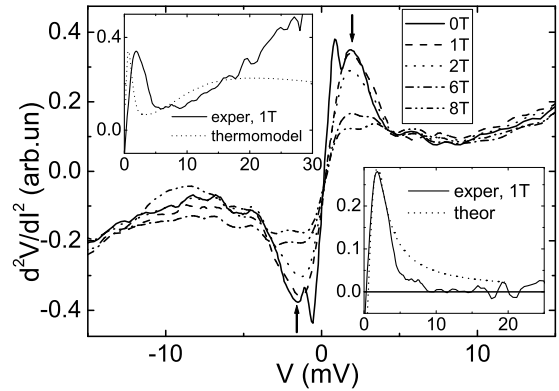}
\end{center}
\caption[] {Experimental spectra ($d^2V/dI^2$) of a
PrOs$_4$Sb$_{12}$--Cu PC from Fig.1. Arrows mark the CEF peak. Top
inset: Zero magnetic field $d^2V/dI^2$ of the same contact at
$B$=1\,T in the wider bias range in comparison with the thermal
regime calculation (see text). Bottom inset: $d^2V/dI^2$ curve at
1T with the subtracted background of the same PC (solid) compared
to the theoretical one (dotted) calculated according to the theory
\cite{Omel}.} \label{second}
\end{figure}


Finally, we have presented the first results of PCS for the
skutterudite  PrOs$_4$Sb$_{12}$. For low bias a pronounced maximum
in the PC spectra is well resolved. This maximum is suppressed,
broadened and splitted by a magnetic field what points out to its
CEF excitation nature. To reinforce this, measurements at lower
temperature and higher magnetic field are desirable.



\begin{thebibliography}{99}
\bibitem{Bauer} E.D. Bauer, et al., Phys. Rev. B {\bf 65} (2002) 100506.

\bibitem{Maple} M.B. Maple, J. Phys. Soc. Jpn., {\bf 74}  (2005) 222.

\bibitem{Naidyuk} Yu. G. Naidyuk and I. K. Yanson, {\it Point Contact Spect\-ro\-scopy},
Springer Series in Solid-State Sciences, Vol.145 (Springer, New
York, 2004).

\bibitem{Omel} I.O. Kulik, et al.,
Sov. J. Low Temp. Phys. {\bf 14} (1988) 82.

\bibitem{Reiffers} M. Reiffers, et al.,
Phys. Rev. Lett. {\bf 62} (1989) 1560.

\bibitem{Goremychkin} E.A. Goremychkin, et al.,
Phys. Rev. Lett. {\bf 93} (2004) 157003.

\end{thebibliography}
\end{document}